\documentclass[pra,twocolumn,showpacs]{revtex4}
\newcommand{\UQ}{ARC Centre of Excellence for Quantum-Atom Optics, 
School of Mathematics and Physics, University of Queensland, QLD 4072, Australia.}
\usepackage{graphicx,psfrag}
\usepackage{bm}
\usepackage{amssymb}
\usepackage{graphicx}
\usepackage{bm}
\usepackage{color}
\usepackage{psfrag}
\usepackage{epstopdf}

\newcommand{\etal}{{\em et al. }}
\newcommand{\e}{\mbox{e}}
\newcommand{\pr}{Phys. Rev. }

\newcommand{\jpa}{J. Phys. A }

\begin{document}
\title{Asymmetric Gaussian steering: when Alice and Bob disagree}

\author{S.~L.~W. Midgley}
\affiliation{\UQ}
\author{A.~J. Ferris}
\affiliation{\UQ}
\author{M.~K. Olsen}
\affiliation{\UQ}
%-----------------------------------------------------------------------
\date{\today}
%------------------------------------------------------------------------

\begin{abstract}

Asymmetric steering is an effect whereby an inseparable bipartite system can be found to be described by
either quantum mechanics or local hidden variable theories depending on which one of Alice or Bob makes
the required measurements. We show that, even with an inseparable bipartite system, situations can arise where
Gaussian measurements on one half are not sufficient to answer the fundamental question of which theory gives
an adequate description and the whole system must be considered. This phenomenon is possible because of
an asymmetry in the definition of the original Einstein-Podolsky-Rosen paradox and in this article we show
theoretically that it may be demonstrated, at least in the case where Alice and Bob can only make Gaussian
measurements, using the intracavity nonlinear coupler.

%Asymmetric steering is an effect whereby an inseparable bipartite system can be found to be either described by quantum mechanics or local hidden variable theories depending on which one of the famous Alice or Bob makes the required measurements. We show that, even with an inseparable bipartite system, situations can arise where Gaussian measurements on one half are not sufficient to answer the fundamental question of which theory gives an adequate description, but that the whole system must be considered. This phenomenon is possible because of an asymmetry in the definition of the original Einstein-Podolsky-Rosen paradox and in this paper we show theoretically that it may be demonstrated, at least in the case where Alice and Bob can only make Gaussian measurements, using the intracavity nonlinear coupler. 

\end{abstract}

  % checked PACS

\maketitle
%=====================================================================================%

\section{Introduction}

The term steering was introduced by Schr\"odinger in 1935~\cite{schr} as a description of the effect predicted by Einstein, Podolsky and Rosen~\cite{epr}, which has become famous as the EPR paradox. In the language now commonly used in quantum information, steering describes the ability of Alice (or Bob) to perform certain measurements on one half of a non-local entangled system and thereby affect
the ensemble of possible states that describe Bob's (or Alice's) system. In a recent work~\cite{steering}, Wiseman \etal showed that violation of the Reid criteria~\cite{reid} for existence of the EPR paradox is a demonstration of steering for Gaussian systems. Wiseman \etal also raised the very interesting question of whether there could be asymmetric states which could be steerable by one party but not the other. In this paper we answer this question in the affirmative for mixed states and Gaussian measurements~\cite{giedke2002,giedke2003}, showing that this effect can potentially be realised using the intracavity nonlinear coupler~\cite{nlc,mko1}. We note here that it remains an open question whether some form of non-Gaussian measurement is possible on this system which could demonstrate symmetric steering, and also that this is a difficult question to give a general answer to in the case of mixed states. 

The possibility of asymmetric steering or EPR arises from an intrinsic asymmetry in the description which is not present in the usual criteria for establishing the inseparability of a bipartite quantum state (as an example, see the Duan-Simon criteria~\cite{Duan,Simon}). Steering is explained in Wiseman \etal as a task requiring two parties, once again called Alice and Bob. Beginning with Alice, she prepares a bipartite state and sends one half to Bob, with the task being to convince Bob that the state is entangled. Bob believes in quantum mechanics but does not trust Alice; he wants to see the entanglement proven. The two make measurements and communicate their results classically. If the correlations between the two sets of measurements can be explained without invoking quantum mechanics, Bob will not be convinced that the state is entangled. Asymmetric steering arises when Alice can convince Bob that the state is entangled, but after a reversal of the roles, Bob cannot convince Alice of the same thing. As both Alice and Bob are both looking at the same overall system, but get different answers, this is an interesting, albeit different, example of the role of the observer in quantum mechanics. Rather than the usual example where the role of the observer is to collapse the wavefunction, in this case each observer must decide whether the bipartite system can be described by a local hidden variable theory and they get conflicting answers despite making similar measurements on the part available to them.

\section{Steering Criteria}

The EPR criteria developed by Reid~\cite{reid} and violated experimentally~\cite{Ou} by Ou \etal are the criteria needed for demonstrating steering in a continuous variable system for Gaussian measurements. Moreover, in a Gaussian system such as we will be treating here, the inferred quadrature variances as defined by Reid are equivalent to the conditional variances which are optimal for systems which exhibit non-Gaussian statistics. That the Reid criteria give the optimal Gaussian measurement has also been shown more recently by Jones \etal~\cite{jones}, who have shown that if the inferred quadrature variances do not violate the appropriate inequality, the system is not steerable by any Gaussian measurements.  The Reid criteria depend on the predictability of observables on one half of the system from measurements made on the other half. In what follows we will label the two halves as $1$ and $2$ rather than continuing with Alice and Bob. Defining the quadrature operators for each mode as $\hat{X}_{i}=\hat{a}_{i}+\hat{a}^{\dag}_{i}$ and $\hat{Y}_{i}=-i(\hat{a}_{i}-\hat{a}^{\dag}_{i})$, $(i=1,2)$ with $\hat{a}_{i}$ being a bosonic annihilation operator, Reid defines inferred quadrature variances 
\begin{eqnarray}
V_{inf}(\hat{X}_{i})&=&V(\hat{X}_{i})-\frac{[V(\hat{X}_{i},\hat{X}_{j})]^{2}}{V(\hat{X}_{j})},
\nonumber \\
V_{inf}(\hat{Y}_{i})&=&V(\hat{Y}_{i})-\frac{[V(\hat{Y}_{i},\hat{Y}_{j})]^{2}}{V(\hat{Y}_{j})},
\label{vinf}
\end{eqnarray}
where $V(\hat{A},\hat{B})=\langle \hat{A}\hat{B} \rangle -\langle \hat{A}\rangle \langle\hat{B} \rangle$. Obviously we are also free to start with part $i$, with the definitions changed by swapping the indices. When $V_{inf}(\hat{X}_{i})V_{inf}(\hat{Y}_{i})<1$, we have an example of $j$ steering $i$ for Gaussian measurements, whereas an observation of $V_{inf}(\hat{X}_{i})V_{inf}(\hat{Y}_{i})\geq 1$ can be explained classically. This therefore means that a manifestation of
\begin{equation}
V_{inf}(\hat{X}_{i})V_{inf}(\hat{Y}_{i}) < 1 \leq V_{inf}(\hat{X}_{j})V_{inf}(\hat{Y}_{j}),
\label{eq:12}
\end{equation}
would be an example of asymmetric steering for the case of Gaussian measurements. We note here that these inequalities were rigorously defined by Reid~\cite{reid} and are the canonical method for demonstrating the EPR paradox in continuous variable systems. 
Besides being of  fundamental interest in quantum theory, asymmetric steering has been suggested as a possible candidate for use in one-way quantum cryptography~\cite{mko2} and could also have potential applications in the field of quantum control.

\section{Physical System}

\begin{figure}[tbhp]
\includegraphics[width=0.5\textwidth]{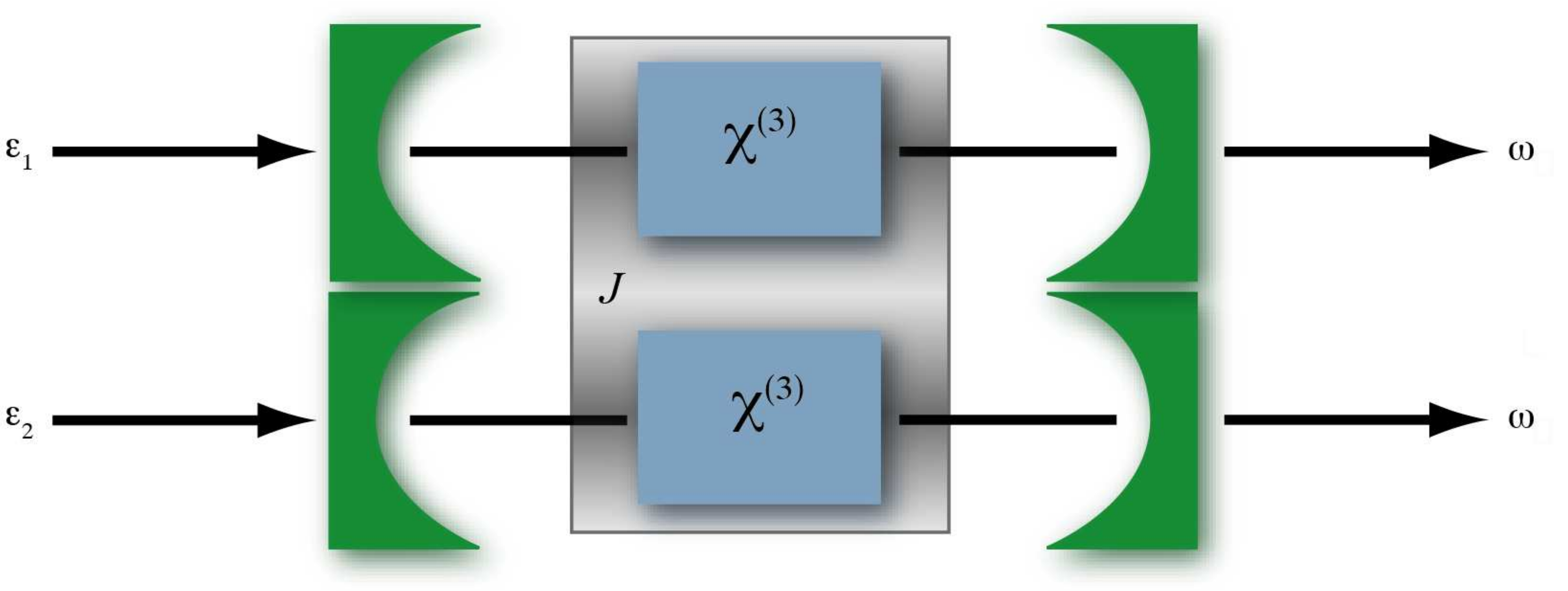}
\caption{(colour online) Schematic of the nonlinear coupler inside pumped coupled Fabry-P\'{e}rot cavities. The two $\chi^{(3)}$ materials act as waveguides with nonlinear interaction strengths $\chi_{i}$ and evanescent coupling strength $J$ are shown, along with the external coherent fields with pump strengths $\epsilon_{i}$, cavity mirrors and optical modes at frequency $\omega$.}
\label{fig:system}
\end{figure}

Having established the inequality which we must satisfy, we now turn our attention to the two-mode system known as the intracavity Kerr nonlinear coupler~\cite{nlc,mko1}. This is a device consisting of two nonlinear $\chi^{(3)}$ media coupled by an evanescent overlap of the guided modes inside optical cavities. A schematic of the device is shown in Fig.~\ref{fig:system}. It is the output beams emerging from the cavities that provide a source of continuous-variable entangled states. In this work we show that asymmetric steering for Gaussian measurements is possible with this device and find parameter regimes for which it is optimised. We find that the effect is large enough that it should be experimentally observable. As far as we are aware, it is not possible to produce this effect via linear optics operations on entangled states.

The Hamiltonian describing the coherently pumped nonlinear coupler may be written as
\begin{equation}
{\cal H} = {\cal H}_{free}+{\cal H}_{int}+{\cal H}_{couple}+{\cal H}_{pump}+{\cal H}_{res},
\label{eq:Hfull}
\end{equation}
where the interaction Hamiltonian is
\begin{equation}
{\cal H}_{int} = \hbar\sum_{i=1}^{2}\chi_{i}\hat{a}_{i}^{\dag 2}\hat{a}_{i}^{2},
\label{eq:Hint}
\end{equation}
the coupling Hamiltonian is
\begin{equation}
{\cal H}_{couple} = \hbar J\left[\hat{a}_{1}\hat{a}_{2}^{\dag}+\hat{a}_{1}^{\dag}\hat{a}_{2}\right],
\label{eq:Hcouple}
\end{equation}
the pumping Hamiltonian is
\begin{equation}
{\cal H}_{pump} = i\hbar\sum_{i=1}^{2}\left[\epsilon_{i}\hat{a}_{i}^{\dag}-\epsilon_{i}^{\ast}\hat{a}_{i}\right],
\label{eq:Hpump}
\end{equation}
\noindent and the reservoir damping Hamiltonian is
\begin{equation}
{\cal H}_{res}= \hbar\sum_{i=1}^{2}\left[\hat{\Gamma}_{i}\hat{a}_{i}^{\dag}+\hat{\Gamma}_{i}^{\dag}\hat{a}_{i}\right],
\label{eq:Hres}
\end{equation}

\noindent where $\hat{a}_{i}$ denote the bosonic annihilation operators in the first and second waveguides for the intracavity modes at frequency $\omega$, the $\chi_{i}$ represents the nonlinear interaction strengths for the two media, the $\epsilon_{i}$ are the classical pumping laser amplitudes, $J$ is the strength of the evanescent coupling, and the $\hat{\Gamma}_{i}$ are the annihilation operators for bath quanta, representing losses through the cavity mirror.
  
In order to describe the intracavity dynamics, we first derive stochastic differential equations (SDEs) in the positive-$P$ representation~\cite{plusP}. This involves using a standard approach~\cite{GardinerQN} whereby we map the quantum operator equations of motion onto a Fokker-Planck equation for the positive-$P$ representation pseudoprobability distribution of the system~\cite{plusP, GardinerQN}, which in turn can be interpreted as a set of $c$-number SDEs. The positive-$P$ method is extremely useful as it allows for the calculation of stochastic trajectory averages which correspond to the normally-ordered expectation values of quantum-mechanical operators.  We define two independent stochastic fields $\alpha_{i}$ and $\alpha^{\star}_{i}$ corresponding to the operators $\hat{a}_{i}$ and $\hat{a}^{\dagger}_{i}$, respectively, in the limit of a large number of stochastic trajectories. Furthermore, we employ the usual zero-temperature Born and Markov approximations \cite{Danbook} when dealing with the reservoir Hamiltonian. The equations of motion are therefore given by,

\begin{eqnarray}
\frac{d\alpha_{1}}{dt}&=&\epsilon_{1}-(\gamma_{1}+i\Delta_{1})\alpha_{1}-2i\chi_{1}\alpha^{+}_{1}\alpha^{2}_{1}-iJ\alpha_{2}\nonumber\\
&+&\sqrt{-2i\chi_{1}\alpha^{2}_{1}}\eta_{1},\nonumber\\
\frac{d\alpha_{1}^{\tiny{+}}}{dt}&=&\epsilon_{1}^{*}-(\gamma_{1}-i\Delta_{1})\alpha_{1}^{+}+2i\chi_{1}\alpha^{+2}_{1}\alpha_{1}+iJ\alpha_{2}^{+}\nonumber\\
&+&\sqrt{2i\chi_{1}\alpha^{+2}_{1}}\eta_{2},\nonumber\\
\frac{d\alpha_{2}}{dt}&=&\epsilon_{2}-(\gamma_{2}+i\Delta_{2})\alpha_{2}-2i\chi_{2}\alpha^{+}_{2}\alpha^{2}_{2}-iJ\alpha_{1}\nonumber\\
&+&\sqrt{2i\chi_{2}\alpha^{2}_{2}}\eta_{3},\nonumber\\
\frac{d\alpha_{2}^{\tiny{+}}}{dt}&=&\epsilon_{2}^{*}-(\gamma_{2}-i\Delta_{2})\alpha_{2}^{+}+2i\chi_{2}\alpha^{+2}_{2}\alpha_{2}+iJ\alpha_{1}^{+}\nonumber\\
&+&\sqrt{-2i\chi_{1}\alpha^{2}_{1}}\eta_{4},
\label{posp}
\end{eqnarray}

%FIX eqn mention
\noindent where $\gamma_{i}$ are the cavity loss rates, and $\eta_{i}$ are real Gaussian noise terms with correlations $\overline{\eta_{i}(t)}=0$ and $\overline{\eta_{i}(t)\eta_{j}(t^{\prime})}=\delta_{ij}\delta(t-t^{\prime})$. We note here that we have included the $\Delta_{j}$ as detunings of the intracavity fields from resonance, where this would have been included in the free Hamiltonian.

\section{Results and discussion}

Using Equation~\ref{posp} we performed dynamical simulations to ascertain the stability of the system for the choice of parameters used in the remainder of this paper. However, as quadrature measurements are usually performed by homodyne measurements in the frequency domain to obtain output spectra, we may perform a much simpler linearized fluctuation analysis of the system to obtain our results. In particular, this analysis of the system allows one to calculate the intracavity squeezing spectra and in turn the output spectral correlations for the cavity \cite{Danbook}. It is these correlations that would be measured in experiments. 

The linearized analysis entails treating the system dynamics as a multivariable Ornstein-Uhlenbeck process, with small and stable fluctuations around steady-state values~\cite{Danbook}. We neglect the noise terms in Equation~\ref{posp} and consider the steady-state solutions of the classical version of the equations of motion. For symmetric parameters as used previously for this system~\cite{mko1}, it was possible to calculate steady-state solutions analytically. In this case, however, we calculated them numerically using a Runge-Kutta method and compared them with the values found from the full positive-P equations, finding excellent agreement. We then proceed to linearize the equations of motion around the steady-state solutions. Specifically, we set $\alpha_{i}=\alpha+\delta\alpha_{i}$, where $\alpha$ is the steady-state mean value and $\delta\alpha_{i}$ are fluctuations, and this gives rise to a set of equations for the fluctuations,

\begin{equation}
d\delta\vec{\alpha}=-A\delta\vec{\alpha} dt +BdW,
\end{equation}

\noindent where $\delta\vec{\alpha}=[\delta\alpha_{1}, \delta\alpha^{+}_{1},\delta\alpha_{2},\delta\alpha^{+}_{2}]^{T}$, $A$ is the drift matrix for the fluctuations, $B$ is the steady-state diffusion matrix and $dW$ is a vector of Wiener increments \cite{GardinerQN}.

The intracavity spectra are then calculated most simply from the Ornstein-Uhlenbeck equations for the system, with the spectral matrix in the stationary state being~\cite{SMCrispin}

\begin{equation}
S(\omega) = \left(A+i\omega\openone\right)^{-1}BB^{T}\left(A^{T}-i\omega\openone\right)^{-1},
\label{eq:spekmat}
\end{equation}

\noindent where $\omega$ is the frequency. Once the intracavity spectra have been calculated, the standard input-output relations~\cite{collett} may be applied in order to arrive at the output spectral variances and covariances. From this we have all the information required to investigate various measures of entanglement, and moreover, study steering in this system. 

To calculate these entanglement measures we need to generalise the definition of the quadrature operators given above to include arbitrary phase angles because, unlike a standard resonant $\chi^{(2)}$ interaction, the Kerr nonlinearity and the detunings both rotate the phase of the fields at which the optimal correlations are found~\cite{mko1,Granja}.
To this end we define the quadrature operator at a given angle $\theta$ as $\hat{X}_{i}^\theta=\hat{a}_{i}e^{-i\theta}+\hat{a}^{\dagger}_{i}\e^{i\theta}$. For compactness, we will use the notation $\hat{X}_{i}^\theta=\hat{X}_{i}$ and $\hat{X}_{i}^{\theta+\frac{\pi}{2}}=\hat{Y}_{i}$, with $\hat{X}_{i}$ and $\hat{Y}_{i}$ obviously being conjugate quadratures with the Heisenberg uncertainty principle requiring $V(\hat{X}_{i})V(\hat{Y}_{i})\geq 1$ . The Reid inequality for steering and EPR can then be written as
\begin{equation}
V_{inf}(\hat{X}_{i})V_{inf}(\hat{Y}_{i}) \geq 1,
\label{eq:reid}
\end{equation}
and holds for arbitrary quadrature angle. To demonstrate asymmetric steering for the case of Gaussian measurements, we need to have $V_{inf}(\hat{X}_{i})V_{inf}(\hat{Y}_{i}) < 1$ for some $\theta$ and $V_{inf}(\hat{X}_{j})V_{inf}(\hat{Y}_{j}) \geq 1$ for all $\theta$. These criteria are exactly equivalent to the criteria used in the work of Jones \etal~\cite{jones}.

As shown in ref.~\cite{mko1}, the intracavity nonlinear coupler demonstrates symmetric steering when the same parameters are used for both sides, that is  $\gamma_{1}=\gamma_{2}$, $\Delta_{1}=\Delta_{2}$, $\epsilon_{1}=\epsilon_{2}$, and $\chi_{1}=\chi_{2}$. To find regimes where asymmetric steering is observable for Gaussian measurements, we introduce asymmetry into the system, with at least one of the conditions $\gamma_{1}\neq \gamma_{2}$, $\Delta_{1}\neq \Delta_{2}$, $\epsilon_{1}\neq\epsilon_{2}$ or $\chi_{1}\neq\chi_{2}$ being true. 

In what follows, we calculate and plot the output spectral Reid correlations, defined in the canonical manner as the appropriate combinations of the Fourier transforms of the two-time quadrature variances and covariances, treated via the standard input-output relations~\cite{collett}. Furthermore, we will use the notation EPR$_{ij}$ to signify the product of the inferred output spectral variances which signify whether $j$ can steer $i$ or not. In order to demonstrate how other commonly used criteria for the detection of entanglement do not detect the asymmetry shown by the steering inequalities, we also calculated the output spectral correlations corresponding to the Duan-Simon criteria~\cite{Duan,Simon} for the same parameters, defined in our notation as
\begin{equation}
V(\hat{X}_{i}\mp\hat{X}_{j})+V(\hat{Y}_{i}\pm\hat{Y}_{j})\geq 4,
\label{eq:Duan}
\end{equation}
where a violation shows that the system density matrix is inseparable. We find that bipartite entanglement is always present as long as one part of the system is steerable.  We note here that the same inequalities, defined above in terms of inferred variances, also hold when written in terms of the the equivalent output spectral correlations. 

\begin{figure}[tbhp]
\includegraphics[width=0.4\textwidth]{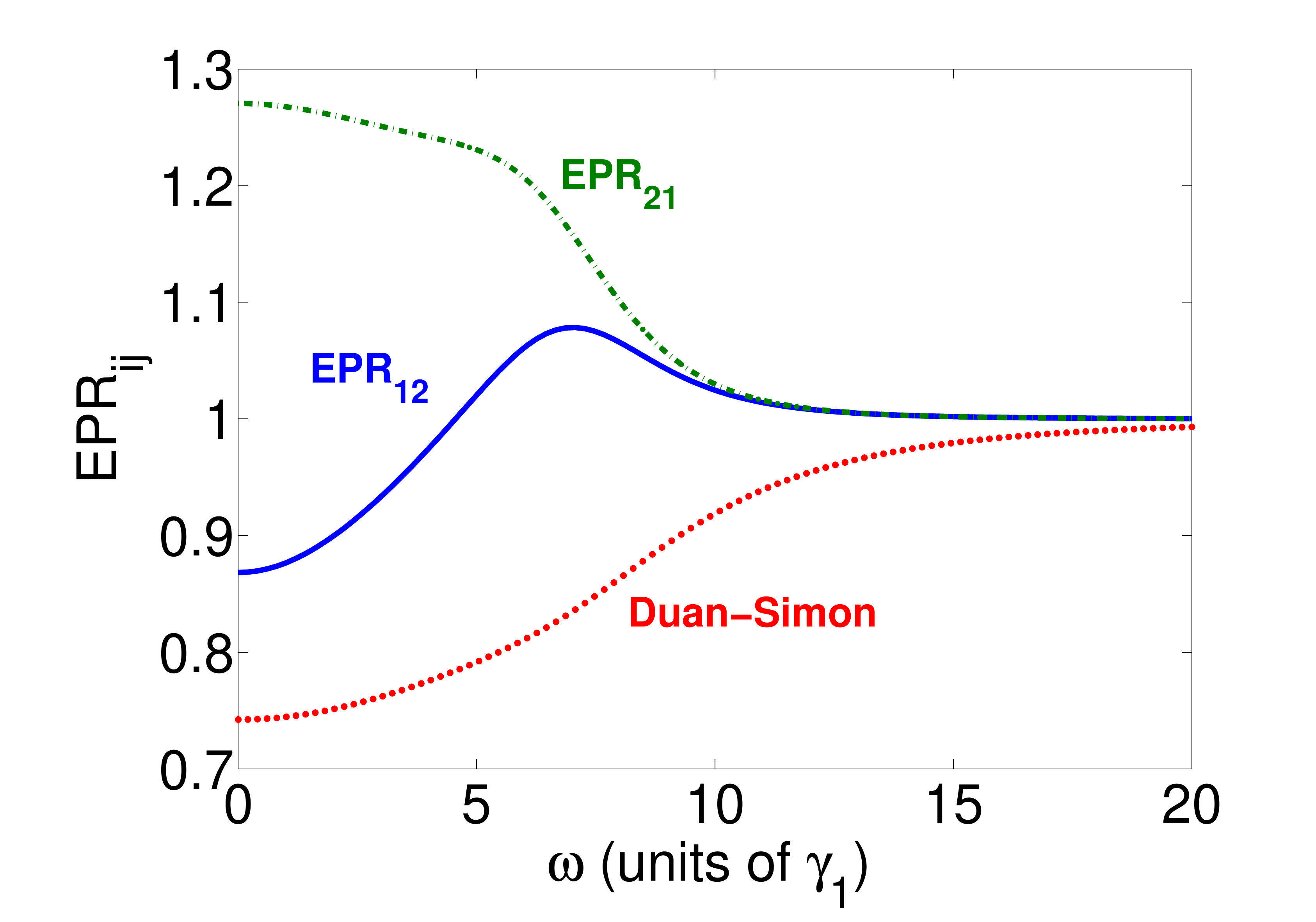}
\caption{(colour online) The output spectral correlations EPR$_{ij}$ for the parameters $\gamma_{1}=1$, $\gamma_{2}=36$, $J=5.0$, $\Delta_{1}=0.001J$, $\Delta_{2}=200\Delta_{1}$, $\epsilon_{1}=10^{3}$, $\epsilon_{2}=80\epsilon_{1}$, $\chi_{1}=10^{-8}$, and $\chi_{2}=10\chi_{1}$. For EPR$_{12}$, $\theta = 9^{o}$, while for EPR$_{21}$, $\theta = 130^{o}$, these angles giving the minimum values. We also show the Duan-Simon correlation for the same parameters, scaled so that a value of less than $1$ signifies bipartite entanglement. All quantities plotted in this and subsequent graphs are dimensionless.} 
\label{fig:best}
\end{figure}

In practice we find that the parameters for each part of the system need to be significantly unbalanced to see a large degree of asymmetric steering for Gaussian measurements. This can be seen in Fig.~\ref{fig:best}, where we show EPR$_{ij}$ in a regime of large asymmetry, and have also plotted a scaled version of the Duan-Simon correlation. For the parameters used, we find that  EPR$_{21}\geq 1$ for all $\omega$ and $\theta$, while EPR$_{12}< 1$ for some choices of $\omega$ and $\theta$. 
Overall, Fig.~\ref{fig:best} represents the signature behavior of asymmetrically steerable states, where Alice can steer Bob's state, but Bob is unable to steer Alice's state. We have therefore demonstrated that asymmetric steering is possible for Gaussian measurements with a relatively simple two-mode system. Furthermore, we show that this can be a substantial effect (with $\sim 40\%$ difference between the values measured on each part), making it amenable to experimental verification. 

\begin{figure}[tbhp]
\includegraphics[width=0.4\textwidth]{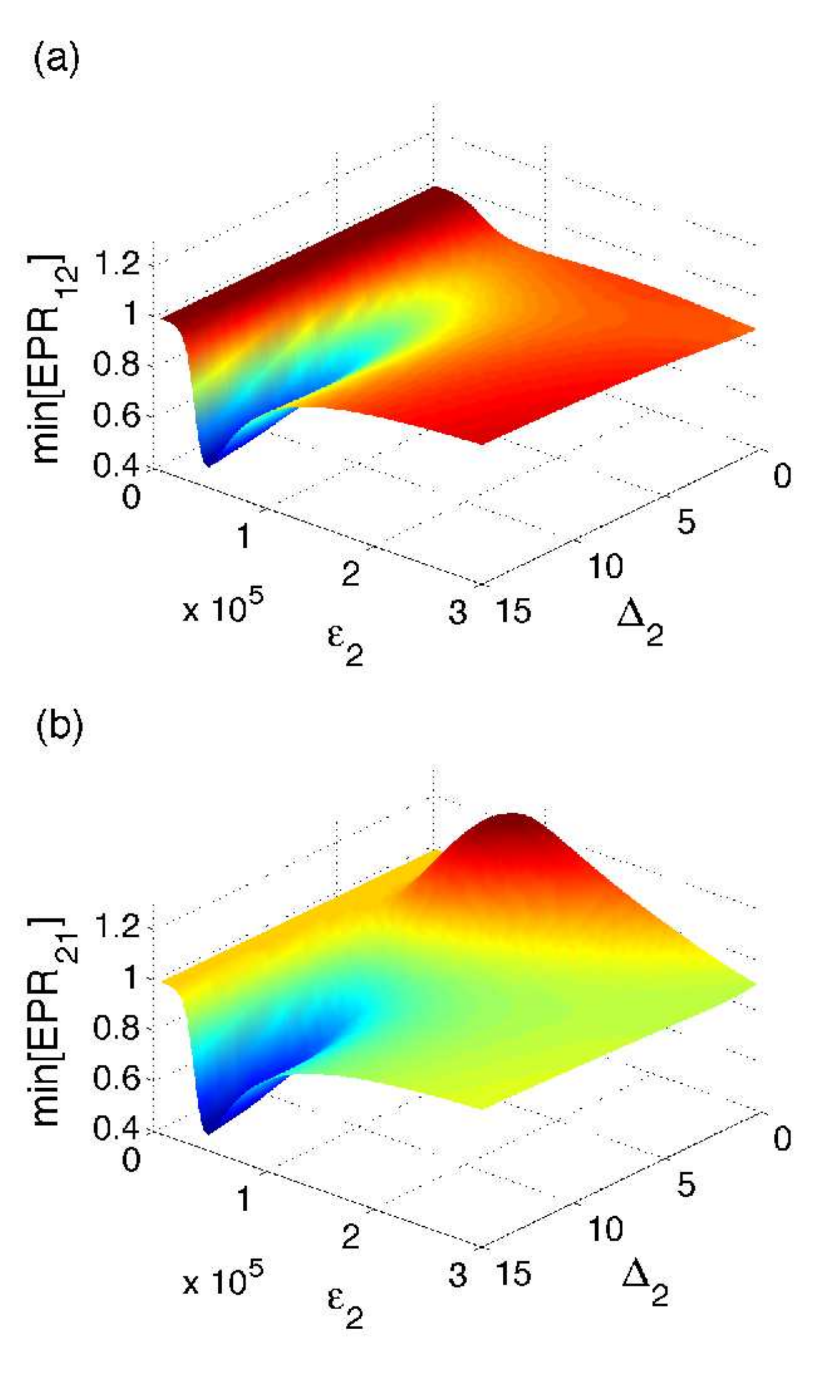}
\caption{(colour online) (a) The minimum of EPR$_{12}$ and (b), the minimum of EPR$_{21}$, as a function of $\epsilon_{2}$ and $\Delta_{2}$, with other parameters the same as in Fig.~\ref{fig:best}. Considering (a) and (b) together, we observe regions of both symmetric and asymmetric steering.}
\label{fig:search1}
\end{figure}

\begin{figure}[tbhp]
\includegraphics[width=0.4\textwidth]{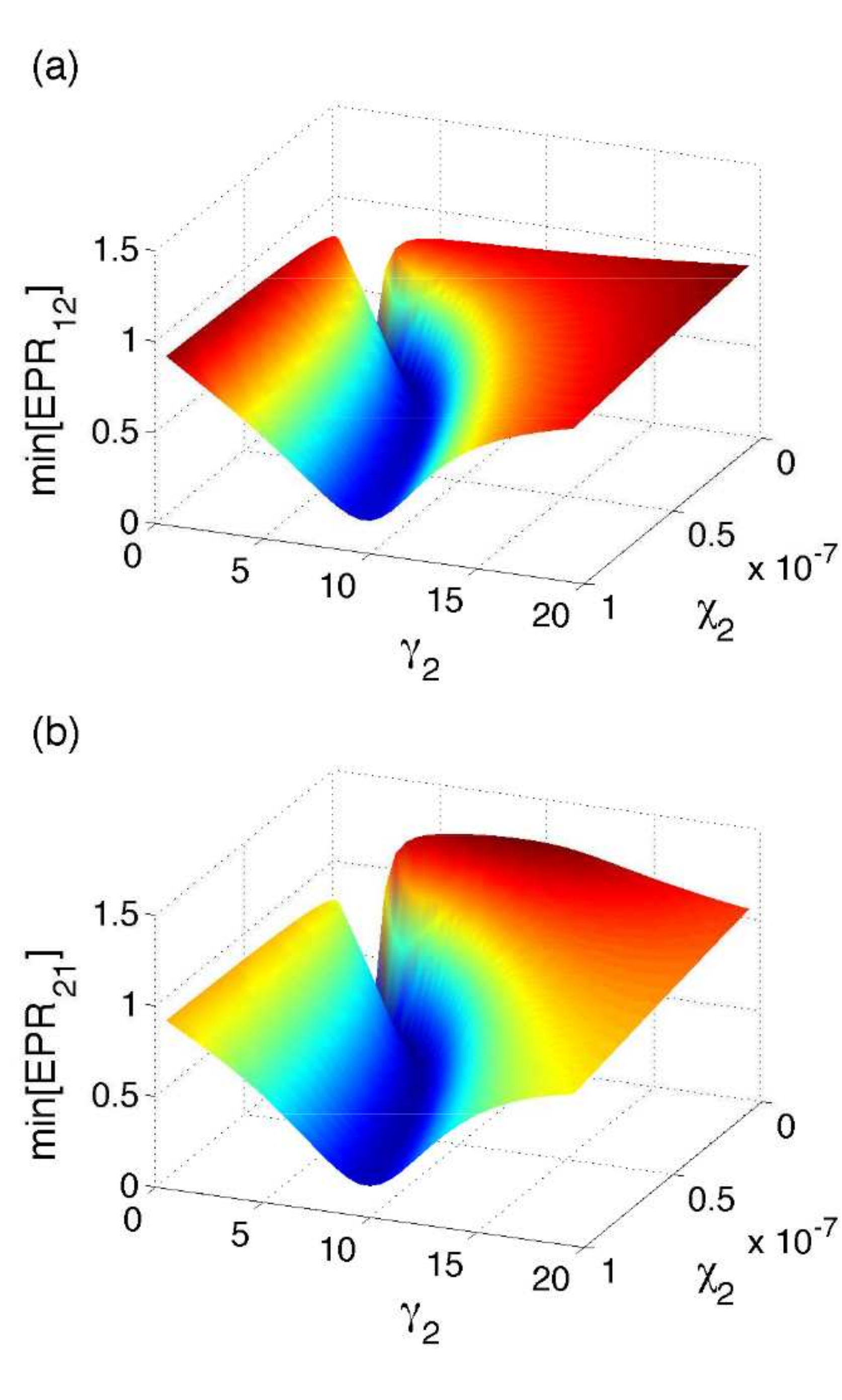}
\caption{(colour online) (a) The minimum of EPR$_{12}$ (b) the minimum of EPR$_{21}$, as a function of $\gamma_{2}$ and $\chi_{2}$, with other parameters the same as in Fig.~\ref{fig:best}. As in Fig~\ref{fig:search1}, we observe both symmetric and asymmetric steering in these plots.}
\label{fig:search2}
\end{figure}

In Fig.~\ref{fig:search1}(a) and (b) we show the EPR$_{ij}$ correlations surrounding the minimum of EPR$_{12}$ and EPR$_{21}$, respectively, as functions of $\epsilon_{2}$ and $\Delta_{2}$, with the mode 1 parameters and the remaining mode 2 parameters fixed. We see that the system exhibits both symmetric and asymmetric behaviour and can observe the transition from one type to the other. For a relatively low pumping rate and a low detuning, we observe the largest degree of asymmetric steering. This is the case shown in Fig.~\ref{fig:best}. On the other hand, symmetric steering is greatest for a large detuning and a low pumping rate.  In Fig.~\ref{fig:search2}(a) and (b), we follow the same procedure, except that we instead vary $\gamma_{2}$ and $\chi_{2}$. In this case, for cavity loss rates in the middle of the range and low nonlinearities the degree of asymmetric steering is at its greatest.

\section{Conclusion}

In conclusion, we have shown that asymmetric steering for the case of Gaussian measurements is possible in a bipartite system and that the intracavity nonlinear coupler is a possible candidate for an experimental demonstration of this extension of the work of Einstein-Podolsky-Rosen and Schr\"odinger. This effect shows that whether a system seems to need a description in terms of quantum mechanics or not can depend on which part is being measured. The important point to note here is that each receiver is allowed to make exactly the same measurements on their part of the inseparable system and obtains contradictory results. This is not a result of one receiver making inappropriate or inaccurate measurements, but is an effect which can only result from the asymmetry present in both the system and the way that steering is defined in terms of the EPR paradox. The standard tests for inseparability of the system are completely unable to detect such an asymmetry and there are no possible Gaussian measurements which will allow this system to be steered in both directions. 
As a final point, we note that this effect may have applications in such areas as quantum control and quantum cryptography.

This work was supported by the Australian Research Council Centre of Excellence for Quantum-Atom Optics. We would like to thank Eric Cavalcanti for useful comments.

%******************************************* 

%------------------------------------------
%=================================================================
\end{document}